\documentclass[11pt,subeqn]{article}
\usepackage[margin=1in]{geometry}                
\usepackage{graphicx}
\usepackage{amssymb}
\usepackage{amsmath}
\usepackage{epstopdf}
\usepackage{setspace}
\usepackage{lineno}
\usepackage{authblk} 
\usepackage{threeparttable}
\usepackage{xcolor,colortbl}
\usepackage{chngpage}
\usepackage{float}

\date{}	                                         

\begin{document}

\newcommand{\T}{\textbf{T}}
\newcommand{\K}{\textbf{K}}
\newcommand{\E}{\textbf{E}}
\newcommand{\A}{\textbf{A}}
\newcommand{\bite}{\textsf{\textbf{R}}}

\title{Epidemiological consequences of an ineffective {\em Bordetella pertussis} vaccine}
\author[1,*]{Benjamin M. Althouse}
\author[1]{Samuel V. Scarpino}
\affil[1]{Santa Fe Institute, Santa Fe, NM}

\maketitle

\let\oldthefootnote\thefootnote
\renewcommand{\thefootnote}{\fnsymbol{footnote}}
\footnotetext[1]{To whom correspondence should be addressed. Email: \texttt{althouse@santafe.edu}}

\let\thefootnote\oldthefootnote

\begin{abstract}
The recent increase in {\em Bordetella pertussis} incidence (whooping cough) presents a challenge to global health.  Recent studies have called into question the effectiveness of acellular {\em B. pertussis} vaccination in reducing transmission. Here we examine the epidemiological consequences of an ineffective {\em B. pertussis} vaccine.  Using a dynamic transmission model, we find that: 1) an ineffective vaccine can account for the observed increase in {\em B. pertussis} incidence; 2) asymptomatic infections can bias surveillance and upset situational awareness of {\em B. pertussis}; and 3) vaccinating individuals in close contact with infants too young to receive vaccine (so called ``cocooning" unvaccinated children) may be ineffective. Our results have important implications for {\em B. pertussis} vaccination policy and paint a complicated picture for achieving herd immunity and possible {\em B. pertussis} eradication.

\end{abstract}

\onehalfspacing


\section*{Introduction}

The worldwide incidence of {\em Bordetella pertussis}, an important causative agent of Whooping Cough, has increased dramatically over the past 20 years and continues to climb~\cite{JacksonRohani2013}.  Last year in the United States alone, there were more diagnosed {\em B. pertussis} cases than in any year since 1955.\footnote{\texttt{http://www.cdc.gov/pertussis/surv-reporting.html}, accessed January 22nd, 2014}  This is despite high vaccination coverage across developed countries~\cite{JacksonRohani2013, Edwards2014, WHOcoverageRates}.  Two general hypotheses have been proposed to explain the rise in {\em B. pertussis} incidence: either vaccination coverage is too low, where individuals remain unvaccinated, or unvaccinated susceptible individuals move into populations; or, vaccinated individuals can still become infected~\cite{JacksonRohani2013, Aguas2006}. While vaccination coverage has likely played a role in increasing incidence, coverage has historically been high ($>90\%$ in many populations)~\cite{JacksonRohani2013,WHOcoverageRates}. This raises the likelihood that the rise in incidence is, at least in part, due to low vaccine efficacy. The increasing {\em B. pertussis} incidence is temporally associated with a change in the vaccine: in the mid-1990s, an acellular vaccine (aP) replaced the highly-effective, but side-effect prone, whole-cell vaccine (wP)~\cite{Edwards2014}. 

There are at least three reasons vaccinated individuals can become infected: one, the vaccine failed to induce sterilizing immunity to the pathogen~\cite{Warfel2014}, two, the vaccine mounted a sterilizing immune response that waned over time~\cite{Wearing2009}, or three, the pathogen evolved to escape sterilizing immunity induced by the vaccine~\cite{Mooi2001}.  A recent study by Warfel, Zimmerman, and Merkel~\cite{Warfel2014} using non-human primates as a model for pertussis infection, suggests that pertussis vaccine efficacy may be more nuanced than previously thought. Their results suggest that individuals vaccinated with current acellular pertussis vaccines (aP) can become asymptomatically infected, and can then transmit infection to susceptible individuals. 

Warfel et al.\ points out that asymptomatic infection of aP vaccinated individuals, and subsequent transmission, may partially account for the increase in observed pertussis incidence. However, from a public health perspective, the presence of vaccinated individuals that can become asymptomatically infected and can transmit disease has profound consequences beyond an increase in incidence. In response to Warfel et al., Domenech de Cell\`es et al. (2014)~\cite{DomenechDeCelles2014} concluded from a qualitative comparison of age-specific infection rates in Sweden and a model of {\em B. pertussis} transmission that aP must protect against transmission. The contrasting experimental and theoretical results on transmission, in the context of potentially life-saving vaccination policy, highlights the need for theoretical and empirical studies of human immunity from aP. 

Using mathematical models of {\em B. pertussis} transmission we explore the effects of asymptomatic transmission by vaccinated individuals on population-level transmission dynamics of pertussis, including how it may render current vaccination policies ineffective. Our results suggest that: one, the use of a non-transmission blocking, or low efficacy vaccine can account for the observed increase in pertussis incidence, two a large asymptomatic infectious class can bias traditional surveillance mechanisms for {\em B. pertussis}, and three, the necessary coverage level for herd-immunity may be unattainable and the practice of ``cocooning" unvaccinated children may be ineffective using only the aP vaccine. 
The results on vaccination have important public health and clinical implications, especially related to recommendations for isolating unvaccinated or partially-vaccinated infants.

\section*{The model}
We formulate a Susceptible, Infected, Removed (SIR) model of pertussis transmission~\cite{Hethcote1997, KeelingRohani2008, AndersonMay1992}. Briefly, susceptible individuals are born at rate $\mu$, where they are vaccinated with whole-cell (wP) or acellular (aP) pertussis vaccine, depending on which vaccine is currently in use.  Our model includes three vaccine epochs: one without vaccination, one with only wP vaccination, and one with only aP. These epochs are non-overlapping, similar to the advent of wP and its eventual replacement by aP~\cite{Edwards2014}. We assume those vaccinated with wP are completely immune to infection (see discussion below).  Those vaccinated with aP move into a vaccinated class where they can become asymptomatically infected. Unvaccinated individuals become infected with pertussis at rate $\beta$ and become symptomatic with probability $\sigma$ (sensitivity to which is explored in the Supplementary Material), and aP vaccinated individuals become asymptomatically infected at rate $\beta$. We assume no difference in transmissibility between symptomatic and asymptomatic individuals (see Supplementary Material). Individuals recover from symptomatic and asymptomatic infection at rates $\gamma_s$ and $\gamma_a$. We assume that aP vaccine is perfectly effective, however this is a conservative assumption with respect to our conclusions. The equations governing transmission dynamics, and analytical expressions for the basic reproduction number and model equilibria are given in the Supplementary Material.

\section*{Can changing to a non-transmission blocking vaccine lead to increased observed pertussis transmission?}

Figure 1 shows the percentage increase in observed symptomatic and unobserved asymptomatic infections after transitioning from a wP to an aP vaccine. As aP vaccination coverage increases, asymptomatic infections increase nearly 30-fold. We see a substantial increase in the observed numbers of symptomatic cases as wP vaccination is replaced by aP vaccination. At low levels of aP vaccination, there is a 5 to 15-fold increase in symptomatic cases. Only at extremely high levels of aP vaccination ($>99\%$) is there no change in symptomatic infections.

\begin{figure}[t]
\begin{center}
\includegraphics[]{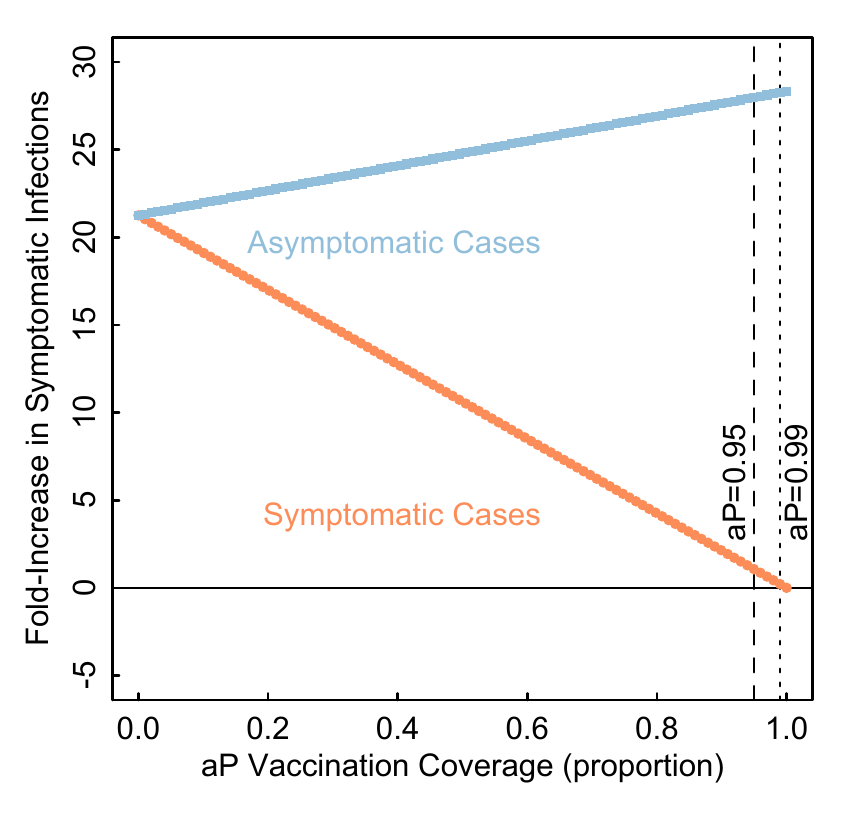}
\caption{\textbf{Can an inefficient vaccine lead to increased transmission?} Figure demonstrates the percentage increase in observed symptomatic and unobserved asymptomatic infections after transitioning from a wP to an aP vaccine, calculated by dividing the steady-state symptomatic cases across varying aP coverage levels with no wP vaccination, to a scenario with 90\% wP coverage and no aP vaccination. Asymptomatic increases were calculated analogously. We see an increase in symptomatic cases across a large range of aP vaccination coverage levels. See Supplementary Information for model details. Parameters: birth rate ($\mu$) = death rate ($\nu$) = 1/75 $\text{years}^{-1}$; recovery rates for symptomatic ($\gamma_s$) and asymptomatic ($\gamma_a$) = 14 $\text{days}^{-1}$; probability of symptomatic infection ($\sigma$) = 0.25; baseline $wP$ vaccination rate = 0.9; transmissibility ($\beta$) is calculated such that $R_0 = 18$.
}
\label{Figure}
\end{center}
\end{figure}

\section*{How does a non-transmission blocking vaccine affect situational awareness?}

From the global to the local-level, accurate knowledge of pertussis incidence (or ``situational awareness") is an essential public health decision support tool, facilitating an assessment of transmission risk, planning for surges in hospitalizations, and in making recommendations for protecting unvaccinated children.  For example, how effective might the strategy of vaccinating individuals in close contact with unprotected children (or ``cocooning"~\cite{Castagnini2012, Healy2011}) be if individuals vaccinated with aP can still transmit disease? Additionally, the mere presence of asymptomatic infections renders an inaccurate assessment of situational awareness through traditional surveillance mechanisms.

Figure 2 demonstrates the percent of the true infections observed at steady-state  ([Observed Incidence/Total Incidence-1]*100) as aP vaccination rate increases and the probability of symptomatic infection ($\sigma$) increases. We find that for realistic aP coverage rates (between 85\% and 95\%), the percentage of total cases expected to be observed is low ($<15\%$), and are highly dependent on the probability of an infection becoming symptomatic (a parameter that is generally not known).  These results are likely to be conservative given the low, but unknown, diagnosis rate of asymptotic infections and known underreporting of symptomatic infections in adults~\cite{RendiWagner2010}.

\begin{figure}[t]
\begin{center}
\includegraphics[]{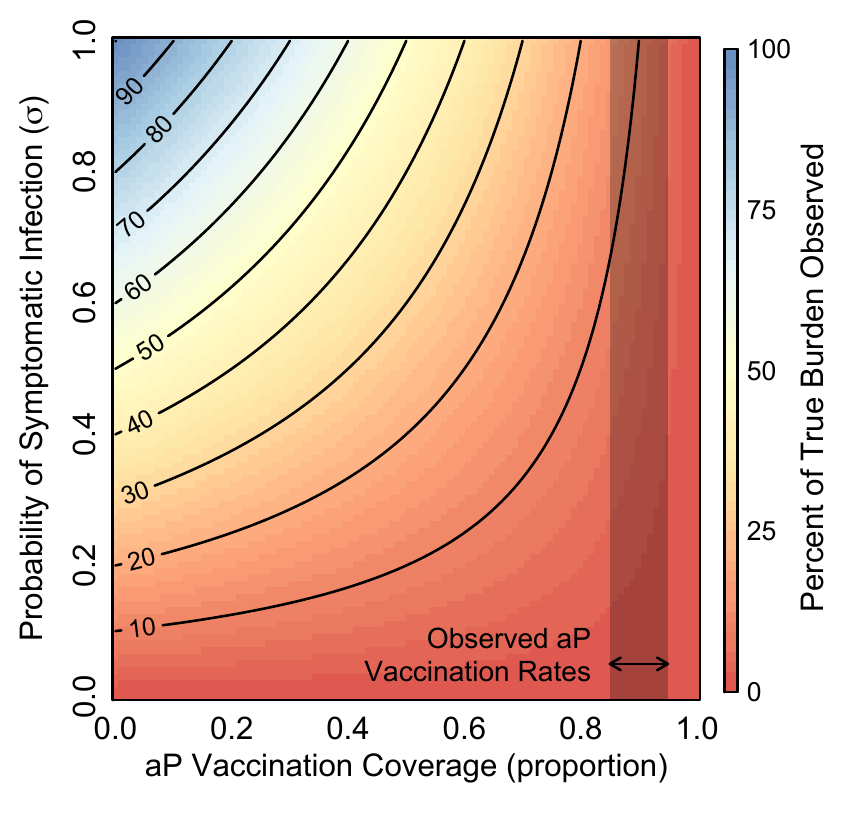}
\caption{\textbf{How does an inefficient vaccine affect situational awareness?} Figure shows the percent difference in observed infections (symptomatic) from true infections (symptomatic + asymptomatic) at steady-state as aP vaccination rate increases and the probability of symptomatic infection increases. Shaded area indicates a range of reasonable aP vaccination rates. At current aP vaccination coverage levels, the majority of cases are asymptomatic and therefore undetected. See Supplementary Information for model details. Parameters: birth rate ($\mu$) = death rate ($\nu$) = 1/75 $\text{years}^{-1}$; recovery rates for symptomatic ($\gamma_s$) and asymptomatic ($\gamma_a$) = 14 $\text{days}^{-1}$; baseline $wP$ vaccination rate = 0.9; transmissibility ($\beta$) is calculated such that $R_0 = 18$.
}
\label{Figure}
\end{center}
\end{figure}

\section*{Effects on herd immunity}
A primary public health objective of vaccination is to achieve herd immunity without exposing individuals to the morbidity and mortality associated with infection~\cite{Fine2011}. Herd immunity is achieved when there is sufficient population-wide immunity to disrupt transmission.  Attaining vaccination-levels high enough for herd immunity (termed the critical vaccination threshold) is a first step towards eradication.  Thus from both an individual perspective and a public health perspective, the level of vaccination needed for herd immunity is an important quantity.  

Figure 3 illustrates the critical aP vaccination threshold. Importantly, this calculation takes into account the coverage level of those individuals previously vaccinated with the transmission-blocking wP vaccine. In ranges of $R_0$ consistent with those observed for pertussis (16-20)~\cite{Anderson1990}, the aP critical vaccination threshold is greater than $95\%$, even with perfect wP vaccination coverage.  Furthermore, with a modest $R_0=5$~\cite{Kretzschmar2010}, the necessary aP vaccination coverage is greater than 80\% (again with perfect wP vaccination coverage). These results have two implications: one, herd immunity may be unattainable and ``cocooning" ineffective with the current aP vaccine and two, given the likelihood of waning immunity and less than perfect vaccine coverage, the window of opportunity to achieve herd immunity, if it exists, is rapidly closing.

\begin{figure}[t]
\begin{center}
\includegraphics[]{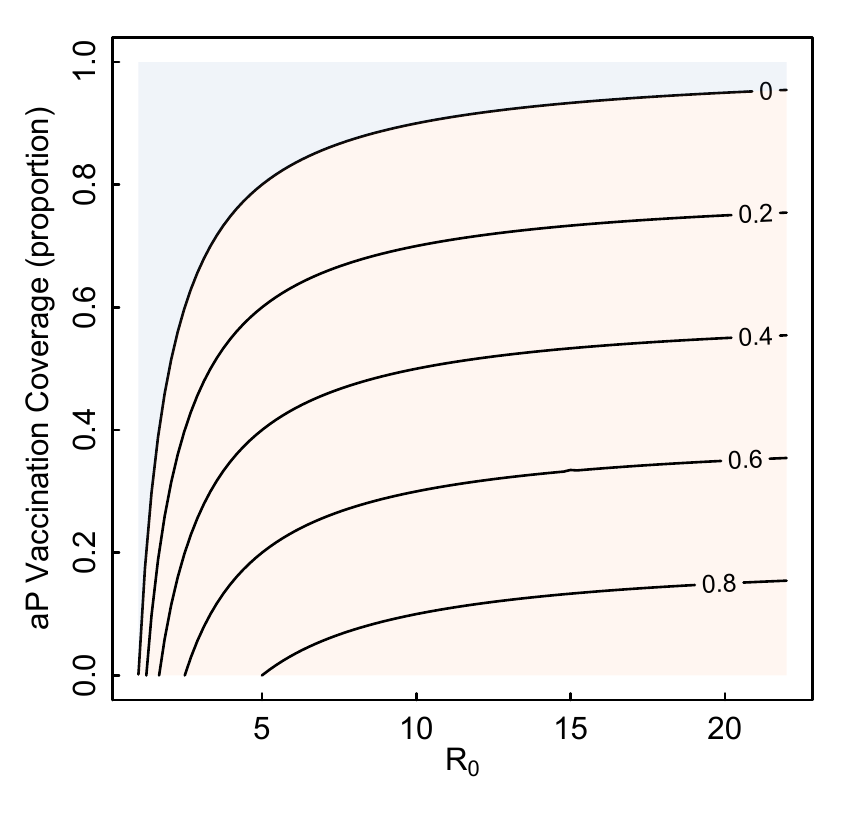}
\caption{\textbf{How does an inefficient vaccine affect herd immunity?} Figure plots the critical aP vaccination threshold required to achieve herd immunity taking into account previous wP vaccination levels (contours). Across a broad range of $R_0$ values for {\em B. pertussis}, herd immunity can only be achieved with high coverage rates of both wP and aP vaccination (blue region). Additionally, we find that as wP vaccination levels increase, less aP coverage is necessary to achieve herd immunity (red region). See Supplementary Information for model details. Parameters: birth rate ($\mu$) = death rate ($\nu$) = 1/75 $\text{years}^{-1}$; recovery rates for symptomatic ($\gamma_s$) and asymptomatic ($\gamma_a$) = 14 $\text{days}^{-1}$; probability of symptomatic infection ($\sigma$) = 0.25; transmissibility ($\beta$) is calculated per value of $R_0$.
}
\label{Figure}
\end{center}
\end{figure}

\section*{Discussion}
An ineffective aP vaccine can account for the increase in pertussis incidence without waning immunity or evolution of pertussis away from protective immunity; complicates situational awareness surrounding levels of current pertussis transmission; and makes achieving herd immunity difficult, if not impossible.  Importantly, if transmission occurs via asymptomatically infectious vaccinated individuals (as suggested in Warfel et al.\ (2014)), the future of global {\em B. pertussis} control could be in jeopardy.
 

As is the case with all models, the one used in this study makes a number of simplifying assumptions.  However, most of these assumptions likely render our conclusions conservative.  We assume wP vaccination is 100\% effective, which may not be the case~\cite{Bentsi1997}. However, this is analogous to having lower coverage overall, and importantly implies that the coverage required for herd immunity will be even higher. Our model does not explicitly account for waning immunity or evolution of the pertussis bacterium~\cite{Wearing2009, Queenan2013} -- two factors which likely play a large role in the epidemiological dynamics of {\em B. pertussis}.
However, inclusion of one or both of these would merely increase the number of individuals susceptible to both symptomatic and asymptomatic infection. Importantly, the qualitative results of this paper will only be exacerbated by waning vaccine immunity and/or the evolution of vaccine resistance. 

Our model also assumes that symptomatic and asymptomatic infections have the same basic reproduction number. Although coughing may increase transmission, the total bacterial load in the nasopharynx of {\em B. pertussis}-infected non-human primates is similar between symptomatic and asymptomatic individuals~\cite{Warfel2014}. The same study suggested that the duration of higher bacterial loads may be longer in asymptomatic individuals. However, and perhaps more importantly, being asymptomatic suggests individuals may not alter their behavior and thus contact more individuals than a symptomatic individual. Therefore, it seems more plausible that the $R_0$ for aP vaccinated individuals is higher and thus have underestimated the critical vaccination threshold for achieving herd immunity.

That there has been a rise in whooping cough incidence is irrefutable. The findings presented in Warfel et al., in conjunction with ours, have profound implications for the understanding of {\em B. pertussis} transmission dynamics and for vaccination policy.  Specifically, our results may explain the negative outcome found in a recent study of postnatal cocooning~\cite{Castagnini2012} and presents a complicated picture for achieving herd immunity and possible eradication. Although we now have conflicting theoretical~\cite{DomenechDeCelles2014}, and experimental results~\cite{Warfel2014} surrounding {\em B. pertussis} transmission, both have serious caveats. The experimental work was done in a non-human primate system that may or may not be a good approximation for human transmission dynamics, and the theoretical work did not present a rigorous evaluation of the transmission hypothesis. 
Clearly more research is necessary, but if our results hold, public health authorities may be facing a situation similar to that of polio and OPV virus transmission~\cite{Fine1999}.  Again suggesting a modification of recommendations to clinicians for protecting unvaccinated children. In light of current evidence and our results, we cannot simply dismiss the potential far-reaching epidemiological consequences of an ineffective {\em B. pertussis} vaccine.

\paragraph{Acknowledgements:}
The authors thank David Dowdy, Damien Caillaud, and Laurent H\'ebert-Dufresne for helpful discussions. This work was supported by the Omidyar Foundation and the Santa Fe Institute.


\clearpage
\pagebreak

\large
\noindent \textbf{\textsf{Supplementary Material for Epidemiological consequences of an ineffective {\em Bordetella pertussis} vaccine}}

\normalsize
\noindent Benjamin M. Althouse, Samuel V. Scarpino

\subsection*{Model Equations}

\noindent We formulate a deterministic Susceptible, Infected, Removed (SIR) model of pertussis transmission~\cite{Hethcote1997, KeelingRohani2008, AndersonMay1992}. Briefly, individuals are born susceptible to pertussis infection at rate $\mu$, where they are vaccinated with whole-cell (wP) or acellular (aP) pertussis vaccine at rates $wP$ and $aP$, respectively. We assume those vaccinated with wP are completely immune to infection, while those vaccinated with aP move into a vaccinated class where they are susceptible to asymptomatic infection. Susceptible individuals become infected with pertussis at rate $\beta$ and become symptomatic at rate $\sigma$. Individuals recover from symptomatic and asymptomatic infection at rates $\gamma_s$ and $\gamma_a$, respectively. Individuals die at rate $\nu$, which we set equal to $\mu$ to keep population size constant. The equations governing transmission dynamics are:
\begin{eqnarray}
S'(t) &=& \mu \cdot (1-wP-aP)-\beta    [I_s(t)+I_a(t)] S(t)  -\nu S(t) \label{sEq}\\
I_s'(t) &=& \beta \sigma [I_s(t)+I_a(t)] S(t)-\gamma_s I_s(t) - \nu I_s(t)\\
I_a'(t) &=& \beta (1-\sigma) [I_s(t)+I_a(t)] S(t)+\beta [I_s(t)+I_a(t)] V(t)-\gamma_a I_a(t) - \nu I_a(t)\\
V'(t) &=& \mu \cdot aP-\beta [I_s(t)+I_a(t)] V(t)- \nu V(t)\\
R'(t) &=& \mu \cdot wP +\gamma_s I_s(t)+\gamma_a I_a(t)- \nu R(t) \label{rEq}
\end{eqnarray}
We begin simulations with neither wP or aP vaccination. After some time period, $t_{wP}$, we initiate wP vaccination, and after that at $t_{aP}$, we stop wP vaccination and begin aP vaccination, similar to replacement of wP by aP vaccines in the mid-1990s~\cite{Edwards2014}. Although this model does not include waning immunity, a process thought to be important for {\em B. pertussis}, we discuss below how this is a conservative modeling choice with respect to our conclusions.

\subsection*{Steady-state Equilibria}

\noindent Calculation of the  stready-state equilibria of this model is done by equating Equations~(\ref{sEq})--(\ref{rEq}) to 0 and solving for the state variables, $S, \ I_s, \ I_a, \ V, \ \text{and} \ R$. There are two equilibria. The disease-free equilibrium is given by:
\begin{eqnarray}
S^* &=& \frac{\mu (1-aP-wP)}{\nu} \label{sEqilibrium}\\
I_s^* &=& I_a^* = 0 \\
V^* &=& \frac{aP \mu}{\nu} \\
R^* &=& \frac{\mu wP}{\nu}. \label{rEqilibrium}
\end{eqnarray}

The other equilibrium has infectious and vaccination classes given by
\begin{eqnarray}
I_s^* &=& -\frac{\sigma \rho \Big[\beta \mu \big[\sigma \rho (\gamma_a-\gamma_s)+\nu (wP-1)+\gamma_s (wP-1)\big]+\nu (\gamma_a+\nu) (\gamma_s+\nu)\Big]}{\beta (\gamma_s+\nu) \big[\sigma \rho (\gamma_a-\gamma_s)+\nu (wP-1)+\gamma_s (wP-1)\big]} \\
I_a^* &=& \frac{(\sigma \rho-wP+1) \Big[\beta \mu \big[\sigma \rho (\gamma_a-\gamma_s)+\nu (wP-1)+\gamma_s (wP-1)\big]+\nu (\gamma_a+\nu) (\gamma_s+\nu)\Big]}{\beta (\gamma_a+\nu) \big[\sigma \rho (\gamma_a-\gamma_s)+\nu (wP-1)+\gamma_s (wP-1)\big]} \\
V^* &=&  -\frac{aP (\gamma_a+\nu) (\gamma_s+\nu)}{\beta \big[\sigma \rho (\gamma_a-\gamma_s)+\nu (wP-1)+\gamma_s (wP-1)\big]}
\end{eqnarray}
where $\rho = (aP+wP-1)$ for clarity.

\subsection*{Calculating $R_0$}

\noindent To calculate the basic reproduction number, $R_0$, we follow the formulation as laid out in Diekmann et al.~\cite{Diekmann2009}. $R_0$ is the spectral radius of the Next Generation Matrix, $\K$, (ie: $R_0 = \rho(\K) = \sup\{\mid \lambda \mid : \lambda \in \sigma(\K)\}$ where $\sigma(\cdot)$ denotes the spectrum of matrix $\K$). We will decompose $\K$ into two matrices: $\T$, the {\em transmission matrix}, where $\T_{ij}$ is the rate at which infected individuals in state $j$ infect individuals in state $i$; and $\Sigma$, the {\em transition matrix}, where $\Sigma_{ij}$ is the rate an individual in state $j$ transitions to state $i$. Diekmann et al. show that 
\begin{equation}
\K = -\E^{T} \T \Sigma^{-1} \E
\end{equation}
Where $\T^{-1}$ is the inverse of matrix $\T$ and $\E$ is a matrix of unit column vectors $\textbf{e}_{ij}$ for all $i$ such that the $i$th row of $\T$ is not identically zero.

We start by linearizing the system about the disease free equilibrium. Assuming $\nu = \mu$, Equations~(\ref{sEqilibrium})-(\ref{rEqilibrium}) become: 
\begin{eqnarray}
S^* &=& (1-aP-wP)\\
I_s^* &=& I_a^* = 0 \\
V^* &=& aP \\
R^* &=& wP.
\end{eqnarray}
 Using the model equations given above, we formulate the {\em infection subsystem} as:
\begin{eqnarray}
I_s' &=& \beta \sigma [I_s+I_a] S^*-\gamma_s I_s - \nu I_s \nonumber \\
&=&  \beta \sigma  [I_s+I_a] (1-aP-wP)-\gamma_s I_s - \nu I_s\\
I_a' &=& \beta (1-\sigma) [I_s+I_a] S^*+\beta [I_s+I_a] V^*-\gamma_a I_a - \nu I_a \nonumber \\ 
&=& \beta (1-\sigma) [I_s+I_a] (1-aP-wP)+\beta [I_s+I_a] aP-\gamma_a I_a - \nu I_a
\end{eqnarray}
where the $(t)$s have been dropped for clarity.

The rate of transmission into symptomatic and asymptomatic classes are given by
\begin{eqnarray}
\frac{\partial}{\partial I_a} (I_s' ) &=& \beta \sigma (1-aP-wP) \\
\frac{\partial}{\partial I_s} (I_a' ) &=& \beta (1-\sigma) (1-aP-wP) \\
\frac{\partial}{\partial V} (I_a' ) &=& \beta aP
\end{eqnarray}

\noindent We find the transmission matrix to be:
\begin{equation}
\T = \left( \begin{array}{ccc}
\beta \sigma (1-aP-wP)& 0 & 0 \\
0 & \beta (1-\sigma) (1-aP-wP) & 0  \\
0& \beta aP & 0 
\end{array} \right).
\end{equation}

Next we calculate the transition matrix, $\Sigma$, where the $(i,j)$ entry is the rate at which an individual in state $j$ transitions to state $i$ (excluding infection transitions). Since there are no transitions between infectious states in our infection subsystem, the transition matrix is a diagonal matrix with the entries equal to the demographic and recovery rates of symptomatics, asymptomatics, and aP vaccinated individuals:

\begin{equation}
\Sigma = \left( \begin{array}{ccc}
-(\nu + \gamma_s) & 0 & 0  \\
0 & -(\nu + \gamma_a) & 0  \\
0 & 0 & \mu aP -\nu
\end{array} \right).
\end{equation}

\noindent This makes finding the inverse of $\Sigma$ trivial, it is simply the reciprocal of each non-zero entry. Now, since none of the rows of $\T$ are identically zero, $\E$ is simply the 4-by-4 identity matrix and our next generation matrix (NGM), is $\K = -\E^{T} \T \Sigma^{-1} \E = -\T  \Sigma^{-1} =$

\begin{equation} \label{r0full}
\left( \begin{array}{ccc}
-\frac{\beta \sigma (1-aP-wP)}{-(\nu + \gamma_s)} & 0 & 0 \\
0 & -\frac{\beta (1- \sigma) (1-aP-wP)}{-(\nu + \gamma_a)}  & 0 \\
0 & -\frac{aP \beta}{-(\nu + \gamma_a)} & 0  \end{array} \right).
\end{equation}

The eigenvalues of $\K$ are 
\begin{equation}
\left\{ 0, \ \frac{\beta  (\sigma -1) (aP+wP-2)}{\gamma_a+\nu }, \ -\frac{\beta  \sigma  (aP+wP-2)}{\gamma_s+\nu } \right\}.
\end{equation}

$R_0$ is defined as the dominant eigenvalue of $\K$, which is determined by the values of the parameters. The second eigenvalue corresponds to the $R_0$ of the asymptomatic strain, and the third to the $R_0$ of the symptomatic strain. Thus, $R_0$ for the entire system is given by the sum:
\begin{eqnarray}
R_0 &=&  \frac{\beta  (\sigma -1) (aP+wP-1)}{\gamma_a+\nu } -\frac{\beta  \sigma  (aP+wP-1)}{\gamma_s+\nu } \\
&=& \frac{ (1 - aP - wP) \beta (\gamma_s + \nu + \gamma_a \sigma - \gamma_s \sigma)}{(\gamma_a+\nu ) (\gamma_s+\nu )}
\end{eqnarray}

\subsection*{Calculating the Herd Immunity Threshold}

To achieve herd immunity, $R_0$ for the system must be less than 1. Thus, if $r_0$ is the basic reproductive number for the system in absence of any vaccination, $R_0 =  (2 - aP - wP) \cdot r_0$, and the critical vaccination threshold for aP vaccine, $aP_{c}$, is
\begin{eqnarray}
R_0 &=&  (1 - aP - wP) r_0 <1 \\
&\implies& aP_{c} < 1 - 1/R_0 - wP
\end{eqnarray}
By symmetry, $wP_{c} < 1 - 1/R_0 - aP$. 

Note that in many scenarios, $aP_{c}>1$, and eradication is impossible by a single vaccination alone.

\subsection*{Different Forces of Infection}

Asymptomatic infection may be less transmissible than symptomatic infection due to less shedding of bacteria through coughing. On the other hand, symptomatic individuals may have a smaller force of infection due to self isolation. Thus, in the main text we assume equal forces of infection. To assess sensitivity of our results to this assumption, we can formulate the model with unequal forces of infection, $\beta_s$ and $\beta_a$, for symptomatic and asymptotic infections, respectively. We modify the infection term
\begin{equation}
\beta (1-\sigma) [I_s(t)+I_a(t)] S(t) \to (1-\sigma) [\beta_s I_s(t)+ \beta_a I_a(t)] S(t)
\end{equation}

\noindent The steady-state equilibrium becomes
\begin{eqnarray}
\textstyle I_s^* &=&\textstyle -\frac{\sigma (aP+wP-2) \left(\beta_a \mu (\gamma_s+\nu) (\sigma (aP+wP-2)-wP+2)-(\gamma_a+\nu) \left(\beta_s \mu \sigma (aP+wP-2)+\nu^2+\gamma_s \nu\right)\right)}{(\gamma_s+\nu) (\beta_a (\gamma_s+\nu) (\sigma (aP+wP-2)-wP+2)-\beta_s \sigma (aP+wP-2) (\gamma_a+\nu))} \\
\textstyle I_a^* &=& \textstyle \frac{(\sigma (aP+wP-2)-wP+2) \left(\beta_a \mu (\gamma_s+\nu) (\sigma (aP+wP-2)-wP+2)-(\gamma_a+\nu) \left(\beta_s \mu \sigma (aP+wP-2)+\nu^2+\gamma_s \nu\right)\right)}{(\gamma_a+\nu) (\beta_a (\gamma_s+\nu) (\sigma (aP+wP-2)-wP+2)-\beta_s \sigma (aP+wP-2) (\gamma_a+\nu))}
 \\
\textstyle V^* &=& \textstyle \frac{aP(\gamma_a+\nu) (\gamma_s+\nu)}{\beta_a (\gamma_s+\nu) (\sigma (aP+wP-2)-wP+2)-\beta_s \sigma (aP+wP-2) (\gamma_a+\nu)}.
\end{eqnarray}

\noindent And $R_0$,
\begin{equation}
R_0  = (2 - aP - wP)\left( \frac{ \beta_a (\sigma - 1)}{(\gamma_a + \nu)} - \frac{\beta_s \sigma}{(\gamma_s+\nu )}\right).
\end{equation}

For a less transmissible asymptomatic infection, none of the results change qualitatively (Figure~\ref{sensDiffBetas}). As $\beta_a$ decreases relative to $\beta_s$, $R_0$ decreases modestly (Figure~\ref{changeR0}).

\subsection*{Sensitivity of Dynamics to Rate of Asymptomatic Infection ($\sigma$)}

Figure~\ref{sensHighSigma} shows the lack of sensitivity of the dynamics to changes in asymptomatic infection rate ($\sigma$).

\noindent\hrulefill\par
\noindent\makebox[\textwidth][c]{%
\begin{minipage}{\linewidth}
\centering
\begin{minipage}[c]{0.45\linewidth}
\begin{figure}[H]
\includegraphics[]{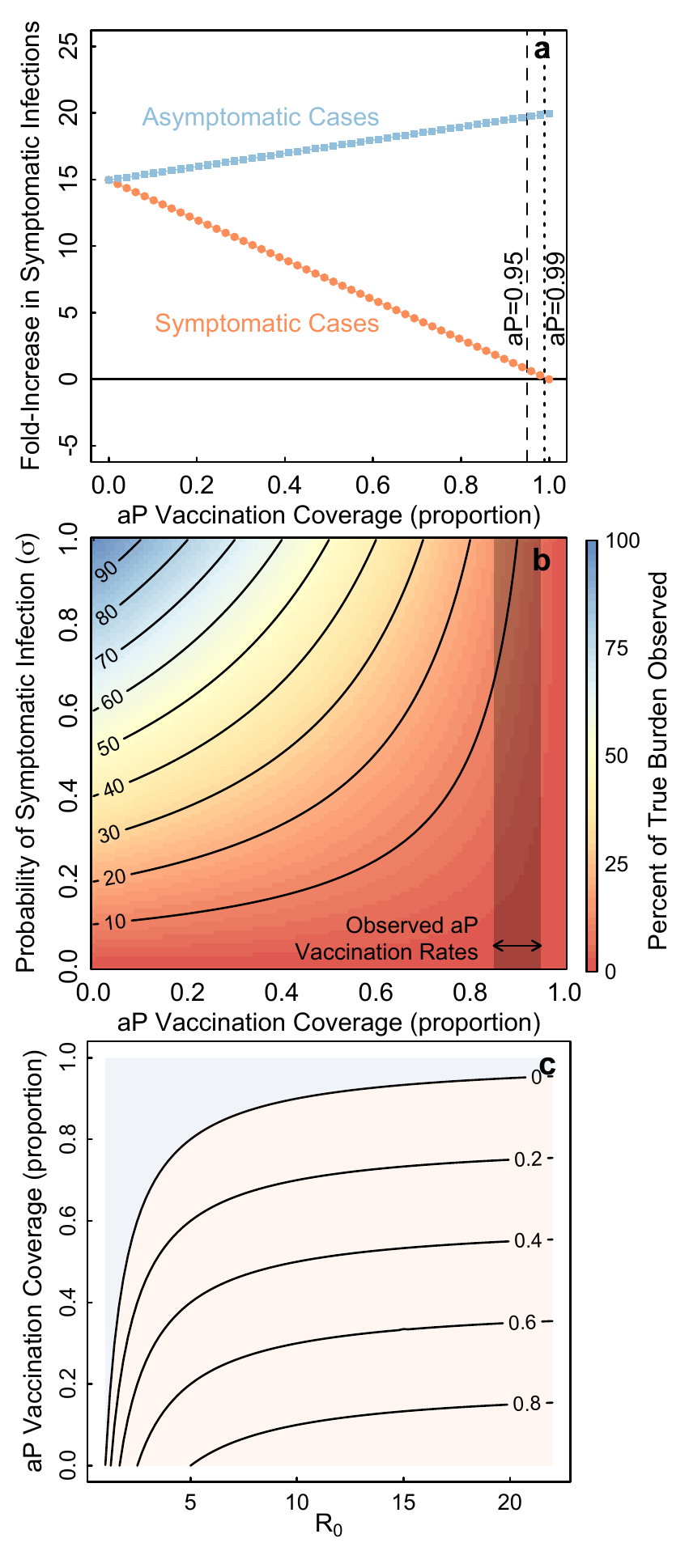}
\caption{\textbf{The effects of ineffective {\em B. pertussis} vaccination under lower asymptomatic transmission} Figure is analogous to the figures in the main text with $\beta_a = \beta_s/10$. Other parameters: $\mu = \nu = 1/75$ $\text{years}^{-1}$;  $\gamma_s = \gamma_a = 14$ $\text{days}^{-1}$; $\sigma = 0.25$; $wP = aP = 0$; $R_0 = 18$.
}
\label{sensDiffBetas}
\end{figure}
\end{minipage}
\hspace{1cm}
\begin{minipage}[c]{0.45\linewidth}
\begin{figure}[H]
\includegraphics[]{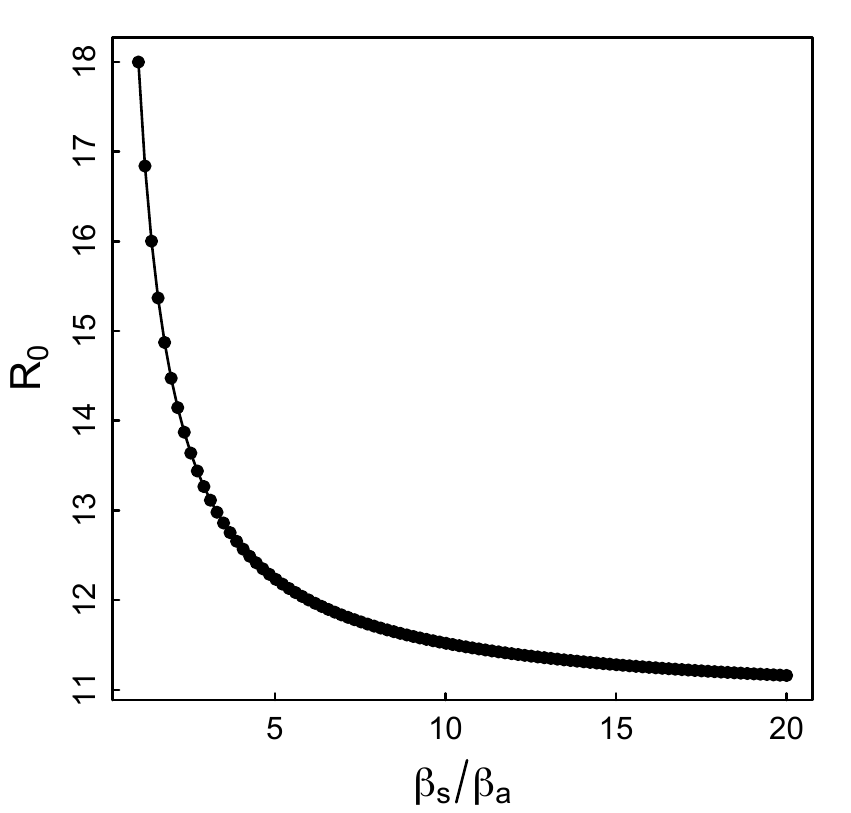}
\caption{\textbf{$R_0$ changes as asymptomatic infections become less transmissible} Figure shows $R_0$ decreasing as asymptomatic infections become less transmissible relative to symptomatic infections ($\beta_s/\beta_a$). Parameters: $\mu = \nu = 1/75$ $\text{years}^{-1}$;  $\gamma_s = \gamma_a = 14$ $\text{days}^{-1}$; $\sigma = 0.25$; $wP = aP = 0$.
}
\label{changeR0}
\end{figure}
\end{minipage}
\end{minipage}
}

\begin{figure}[!t]
\begin{center}
\includegraphics[]{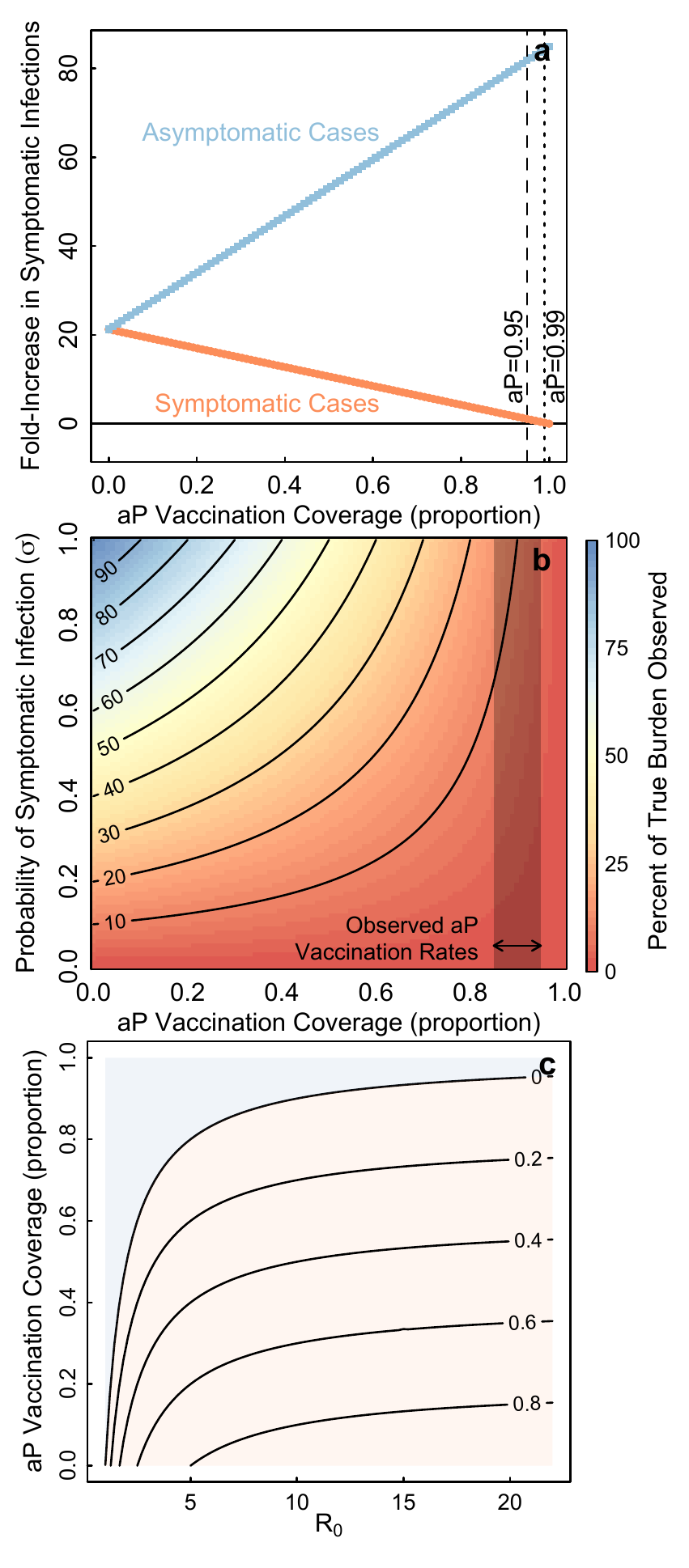}
\caption{\textbf{The effects of ineffective {\em B. pertussis} vaccination under higher asymptomatic infection rate} Figure is analogous to the figures in the main text with $\sigma = 0.75$. Other parameters: $\mu = \nu = 1/75$ $\text{years}^{-1}$;  $\gamma_s = \gamma_a = 14$ $\text{days}^{-1}$; $\sigma = 0.25$; $wP = aP = 0$; $R_0 = 18$.
}
\label{sensHighSigma}
\end{center}
\end{figure}

\pagebreak
\clearpage
\singlespace
\small
\bibliographystyle{ieeetr}
\bibliography{pertussis}

\end{document}